# Speed-of-sound compensated photoacoustic tomography for accurate imaging

**Jithin Jose,[1,†] Rene G. H. Willemink,[2,†] Wiendelt Steenbergen,[1] C. H. Slump,[2] Ton G. van Leeuwen,[1,3] and Srirang Manohar[1]***

[1] *Biomedical Photonic Imaging Group, MIRA-Institute for Biomedical Technology and Technical Medicine, University of Twente, P.O. Box 217, 7500 AE Enschede, The Netherlands*

[2] *Signals and Systems Group, MIRA-Institute for Biomedical Technology and Technical Medicine, University of Twente, P.O. Box 217, 7500 AE, Enschede, The Netherlands*

[3] *Biomedical Engineering and Physics, University of Amsterdam, Academic Medical Center, P.O. Box 22700,1100 DE, Amsterdam, The Netherlands*

[†]Equal contribution

[*]S.Manohar@utwente.nl

**Abstract:**

**Purpose:** In most photoacoustic (PA) tomographic reconstructions, variations in speed-of-sound (SOS) of the subject are neglected under the assumption of acoustic homogeneity. Biological tissue with spatially heterogeneous SOS cannot be accurately reconstructed under this assumption. We present experimental and image reconstruction methods with which 2-D SOS distributions can be accurately acquired and reconstructed, and with which the SOS map can be used subsequently to reconstruct highly accurate PA tomograms.

**Methods:** We begin with a 2-D iterative reconstruction approach in an ultrasound transmission tomography (UTT) setting, which uses ray refracted paths instead of straight ray paths to recover accurate SOS images of the subject. Subsequently, we use the SOS distribution in a new 2-D iterative PA reconstruction approach, where refraction of rays originating from PA sources is accounted for in accurately retrieving the distribution of these sources. Both the SOS reconstruction and SOS-compensated PA reconstruction methods utilize the Eikonal equation to model acoustic wavefront propagation. The equation is solved using a high accuracy fast marching method (HAFMM). We validate the new reconstruction algorithms using numerical phantoms. For experiments we use the recently introduced PER-PACT method which can be used to simultaneously acquire SOS and PA data from subjects. We test the reconstruction algorithms using experimental data acquired with the PER-PACT setup from challenging physical phantoms.

**Results:** It is firstly confirmed that it is important to take SOS inhomogeneities into account in high resolution PA tomography. The iterative reconstruction algorithms, that model acoustic refractive effects, in reconstructing SOS distributions, and subsequently using these distributions to correct PA tomograms, yield artifact-free highly accurate images. Our approach of using the hybrid measurement method and the new reconstruction algorithms, is successful in substantially improving the quality of PA images with a minimization of blurring and artefacts.

## 1: Introduction

In recent years, photoacoustics (PA) has gained considerable attention in the field of biomedical imaging,[1-5] due to its unique ability to combine the advantages of optical imaging and ultrasound imaging into a single modality. The method is based on generation of acoustic signals from absorbing structures of interest, upon irradiation by pulsed laser light. Since ultrasound scattering in biological tissues is 2-3 orders of magnitude weaker than optical scattering, photoacoustic (PA) imaging provides better resolution than optical imaging for depths greater than 1 mm.[1,3,5] The method has been proven to have great potential in breast cancer imaging,[6-9] melanoma visualization,[10-11] small animal imaging[12-15] and the detection of rheumatoid arthritis[16-17] etc.

An assumption made in most PA image reconstruction algorithms[18-20] is that tissue is homogenous so that a single speed-of-sound (SOS) value for the entire region of interest is applied. This assumption is contrary to the inhomogeneous nature of tissue where, depending on composition, the SOS can vary between 1350 m/s to 1700 m/s.[21] Discrepancies in SOS values can result in reduced image contrast and resolution, and in severe cases also image distortions and artifacts. The problem is somewhat mitigated by manually optimizing an initial guessed average SOS value to maximize sharpness and contrast of certain sought-after features in the image. Such an approach remains an approximate correction which can yield a far from optimal image quality.

Several studies have considered compensation for SOS inhomogeneities in the imaging area in PA tomography[22-26] Xu and Wang[22] were the first to study the effect of acoustic heterogeneities on image reconstruction, and applied a two-layer model for simulating the situation in the breast. They demonstrated that SOS variations led to degradation of images which could be corrected by using an iterative reconstruction method that incorporates complete or partial information of acoustic velocity distribution. Anastasio *et al*[27] demonstrated reconstructions assuming an *a priori* known SOS map, where the PA measurement function was modified by specifying integrations over curved iso-time of flight (TOF) contours instead of circles, where these contours were calculated considering straight ray propagation of ultrasound. Chi Zhang and Wang[25] proposed a method which does not require the knowledge of the SOS distribution. In this method, curved iso-TOF contours were calculated from the correlation between integrated photoacoustic signals from origin-symmetric detector pairs. In the derivation of this method, several approximations are applied such as that the imaged object is small or equivalently that the detectors are far away from the imaged object and further that the SOS inhomogeneities are not too high. Recently, Treeby *et al*[26] proposed an autofocus approach for automatically selecting the SOS. The approach was based on maximizing the sharpness of the reconstructed image as quantified by a focus function.[26]

A first experimental approach to map the SOS distribution in a PA setting was proposed by Jin and Wang,[28] in the form of ultrasound transmission tomography (UTT).[29-30] They introduced an extra ultrasound transmitter in a conventional PA imaging setup. Reconstruction involved integrations over iso-TOF contours, with these contours calculated using straight ray propagation paths. A drawback of the method is that it requires an extra ultrasound transmitter and an additional measurement to obtain the SOS map.

Recently, we proposed a hybrid method to obtain the acoustic property distributions simultaneous with PA tomography.[31-33] The method, called passive element enriched photoacoustic computed tomography (PER-PACT), is based on the use of 'passive' elements of ultrasound. We define 'passive' element as a strong absorber of light which functions as an ultrasound generator by the PA effect, when placed in the path of the laser pulse. The situation is in contrast to conventional electrical activation of piezoelectric (active) elements for producing ultrasound. The laser-induced ultrasound transient created in the passive element is used to perform UTT which permits the development of 2-D slices of SOS and acoustic attenuation (AA) distribution in the object. Further, the passive element is chosen to possess a small geometrical cross-section to intercept only a small part of the illuminating beam; most of the light passes on to illuminate the object so that simultaneous PA tomography of the object can be performed. We have earlier shown the feasibility of this hybrid tomography approach using phantoms and biological tissues.[32] We further modified the approach introducing multiple passive elements to reduce imaging time without compromising image quality.[34] On the reconstruction side, we proposed an algorithm for reconstruction of PA image taking SOS spatial variations into account and tested this on numerical phantoms.[33]

In this paper, we revisit the 2-D reconstruction method which accounts for the actual SOS distribution in reconstructing the PA image, and validate it with both simulated and experimental data. For the latter, we use the hybrid method to acquire data from phantoms and biological specimens which possess acoustic property and optical absorption inhomogeneities. The specific advancements here compared with earlier work[32,33] may be consolidated as follows: 1) for retrieving SOS images, an iterative reconstruction algorithm (using the Eikonal equation to model acoustic wave propagation in the forward projection) is used unlike the single-step reconstruction which uses a ray-driven measurement model in Ref. 32; 2) for utilizing SOS information to compensate PA images, a novel iterative PA reconstruction algorithm is used which accounts for ray-refraction also using the Eikonal equation, while previous work[32] used backprojection without SOS compensation, 3) Total Variation (TV) regularization has been used to stabilize the inversion of the matrix of unknown acoustic or optical distributions during reconstruction as an improvement from the previously used gradient-based regularization.[32,33]

## 2: Instrumentation, phantoms and measurement protocol

### 2.1: *Passive element enriched photoacoustic computed tomography (PER-PACT)*

A schematic of the PER-PACT system is shown in Fig. 1. The object under investigation is mounted on a rotary stage and immersed in water in an imaging tank. A Q-switched Nd:YAG laser (Brilliant B, Quantel) operating at 10 Hz repetition rate, delivering 5 ns pulses at a wavelength of 1064 nm is used as light source. The beam is split to illuminate the object from three directions separated by $120^0$. A curvilinear ultrasound detector array (Imasonic, Besançon, France) comprising 32 piezocomposite elements with a central frequency of 6.25 MHz and -6dB frequency bandwidth of 80% is used. The detector has a radius of curvature of 40 mm. Individual elements have dimensions of 10x0.25 mm with an inter-element spacing of 1.85 mm. A 32-channel pulse-receiver system (Lecoeur-Electronique, Chuelles) sampling at 40 $MSs^{-1}$is used for data acquisition.

The array detector was designed to acquire 2-D information in slices through the object. Each detector element is shaped to provide an elevation focus (-3-dB beamwidth of 1 mm) at a distance of 48 mm. The elevation focus possesses a depth-of-focus of approximately 20 mm. When multiple projections of the object have been acquired, we can define an imaging slice of roughly 20-25 mm diameter. All phantoms and objects used (see further Sec. 2.2) have ROIs not exceeding 26 mm in diameter. This justifies the use of the 2-D models used for parameter estimation and image reconstructions (see further Sec. 3).

The passive element comprising a strand of horsetail hair with a diameter of 250 $\mu$m is fixed antipodal to the ultrasound transducer. Multiple projections of the object in the conventional PA sense and in an ultrasound transmission sense are obtained by acquiring ultrasound signals from each detector element for each angle as the object is rotated. The projections permit reconstruction of cross-sectional maps of light absorption and SOS distribution. A description of signal processing, estimation and reconstruction of acoustic transmission parameters are described elsewhere.[32-33] Prior to imaging, a calibration measurement is performed using a reference object (agar gel cylinder carrying a certain number of absorbing landmarks such as horsetail hairs), to ascertain the imaging geometry such as the center of rotation and position of detector elements. Further, PA signals from the samples were deconvolved with signals from a PA point source according to the method of Wang *et al*.[35] The PA point source was created by illuminating a carbon fiber of 10-20 $\mu$m with a sheet of pulsed light.

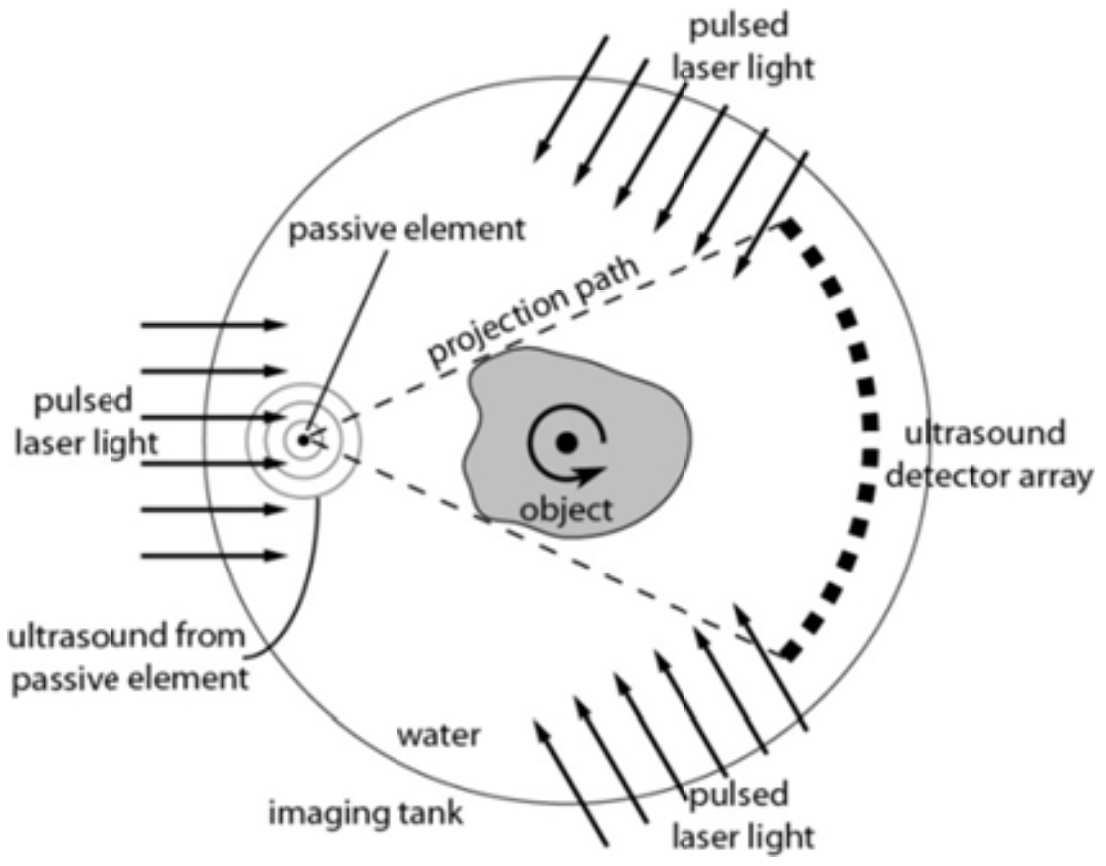

**FIG 1**: Schematic of the experimental setup, with multiple side illumination

### 2.2: *Phantoms used*

To demonstrate the distortion and blurring caused by the assumption of a homogeneous SOS distribution for reconstruction, and to evaluate the performance of the SOS correction algorithm, the following phantoms were used:

(a) *Numerical phantom*: Figures 2(a) and (b) display the SOS and optical absorption distributions in the numerical phantom respectively. The phantom of 26 mm diameter consists of four acoustic inhomogeneities with SOS varying from 1440 $ms^{-1}$ to 1580 $ms^{-1}$ in a background of 1500 $ms^{-1}$. The phantom also possesses five circular optical inhomogeneities structures, with different absorptions and different diameters.

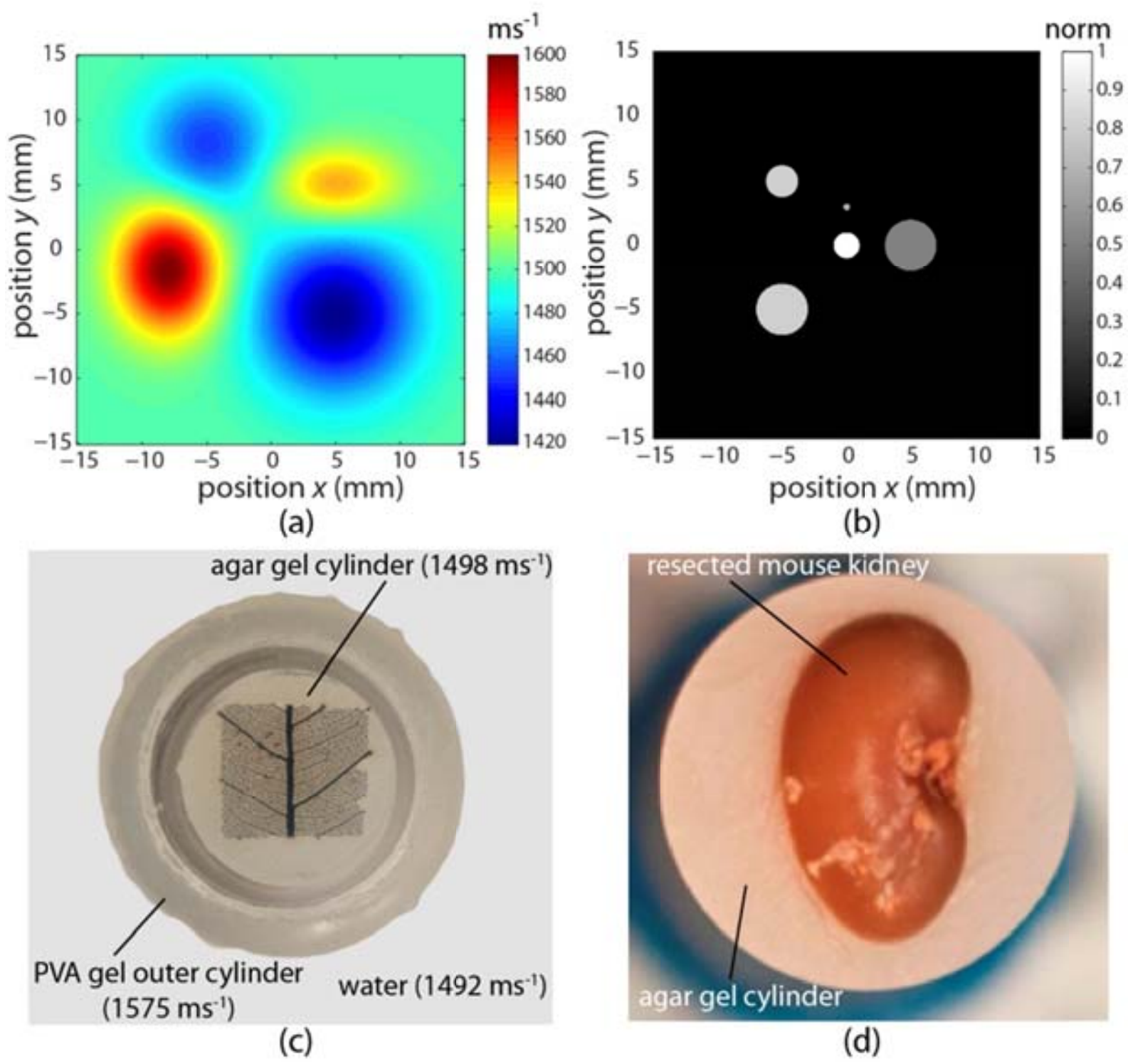

**FIG 2**: Overview of the phantoms used. (a) Numerical phantom showing speed-of-sound (SOS) distribution in $ms^{-1}$, (b) same phantom showing the distribution of optical absorption in arbitrary units, (c) photograph of the leaf skeleton piece embedded in an agar cylinder with an enclosing high velocity PVA cylinder. The values of the SOS of various media are also marked. (b) Resected rat kidney embedded in an agar cylinder.

(b) *Leaf skeleton phantom*: A skeletonized leaf from a florist was used to simulate a complex vascular structure. Figure 2(c) shows the top view of the leaf embedded in an agar cylinder with diameter of 26 mm. The dimensions of the piece of leaf skeleton used were approximately 18x18 mm and the sizes of the veins ranged between 1 mm for the midvein down to 30 $\mu$m for quaternary veins. The agar cylinder with the leaf was held within a poly(vinyl alcohol) (PVA) cylinder with inner and outer diameters of 18 mm and 24 mm respectively. PVA possesses an SOS of 1575 $ms^{-1}$ [32, 36] and functions as an SOS inhomogeneity. Eighteen projections of the object were acquired for a slice using a fluence rate of 8 $mJcm^{-2}$ per pulse. Signals were averaged 100 times and the measurement lasted 3 minutes.

(c) *Rat kidney specimen*: For a biological specimen, we used a kidney resected from a freshly sacrificed Wistar rat. The organ was washed three times in PBS (phosphate buffered saline) solution to remove excess blood and then embedded in an agar cylinder of 26 mm diameter. Figure 2(d) shows a photograph of the sample. The coronal plane of the kidney was imaged at a room temperature (22°C) with a fluence rate of 10 $mJcm^{-2}$. Eighteen projections of the PA signals were obtained in a slice and the signals were averaged 100 times.

## 3: Theoretical basis and reconstruction methods

### 3.1: *Photoacoustic wave propagation*

The generation and propagation of photoacoustic pressure waves in acoustically non-attenuating but inhomogeneous SOS media is governed by the PDE[37-38]

$$\nabla^2 p(\mathbf{r},t) - \frac{1}{c^2(\mathbf{r})}\frac{\partial^2 p(\mathbf{r},t)}{\partial t^2} = -\frac{\beta}{C_p}\frac{\partial I(t)}{\partial t}A(\mathbf{r}),$$

(1)

where $p(r,t)$ is the generated pressure at location $r$ and time $t$, $\beta$ is the volume thermal expansion coefficient, $C_p$ is the specific heat, $c(r)$ is the acoustic speed distribution, $I(t)$ is the laser pulse profile, and $A(r)$ the optical absorption distribution. In the case of stress confinement, for the inhomogeneous SOS distribution case, an approximate solution to Eq. (1) in 2-D, was found by Jin and Wang as[28]

$$p(\mathbf{r},t) = \eta\frac{\partial}{\partial t}\iint\limits_{t=t_f(\mathbf{r}',\mathbf{r})} \frac{A(\mathbf{r}')}{|\mathbf{r}-\mathbf{r}'|}d\mathbf{r}',$$

(2)

where $\eta$ is a constant, and $t_f(\mathbf{r}',\mathbf{r})$ is the TOF for a pressure wave to travel from point $\mathbf{r}$ to point $\mathbf{r}'$. Equation (2) shows that the generated pressure can be seen as the projections over iso-TOF contours determined by function $t_f(\mathbf{r}',\mathbf{r})$, which depends on the SOS distribution $c(\mathbf{r})$. From this relation, it is observed that given an SOS distribution, the relation between optical absorption $A(\mathbf{r})$ and $p(\mathbf{r},t)$ is linear.

We outline here the procedure to reconstruct $A(\mathbf{r})$ from a given number of $p(\mathbf{r},t)$ acquired in the time-domain with a certain sampling frequency, at different spatial positions $\{\mathbf{r_1},\dots,\mathbf{r_n}\}$ external to the object. We hold the measurement vectors $\mathbf{z}_{\mathrm{p,i}}$, containing the sampled time-domain signals $p(\mathbf{r_i},t)$ at the various $\mathbf{r}_{\mathrm{i}}$ in a complete measurement vector $\mathbf{z}_{\mathrm{p}} = [\mathbf{z}_{\mathrm{p,1}}^T,\dots,\mathbf{z}_{\mathrm{p,n}}^T]^T$. The continuous $A(\mathbf{r})$ is discretized onto a uniformly sampled grid and stored in vector $\boldsymbol{x}_{\mathrm{A}}$. If we know for every detector position, the corresponding TOF values at each of the spatial grid coordinates of $\boldsymbol{x}_{\mathrm{A}}$, the forward model can be written as a specific sum of elements in $\boldsymbol{x}_{\mathrm{A}}$ followed by a differentiation operation. In matrix notation, this is expressed as:

$$\mathbf{z}_{\mathrm{p}} = \mathbf{H}_{\mathrm{dt}}\mathbf{H}_{\mathrm{TOF}}\boldsymbol{x}_A$$

(3)

where $\mathbf{H}_{\mathrm{dt}}$ is a large and sparse matrix representing the time derivative, and $\mathbf{H}_{\mathrm{TOF}}$ is a large and sparse matrix representing the integral (compare with Eq. 2). To solve for $\boldsymbol{x}_{\mathrm{A}}$, we employ algebraic reconstruction in an iterative framework. We discuss this in detail further in Sec. 3.3, but first we present our approach to reconstruct the SOS distribution and subsequently $\mathbf{H}_{\mathrm{TOF}}$.

### *3.2: Reconstruction of SOS distribution incorporating refraction of rays*

*Forward projector*

In iterative reconstructions, it is crucial to image quality that the ray integrals closely model the physical situation of energy propagation. The higher the fidelity of the forward projector, the more accurate are the corrections computed by comparing the integrals of the current reconstruction estimate with the acquired data.[39] For the forward projector we utilize the Eikonal equation which can be used to model acoustic wavefront propagation with inhomogeneous SOS distributions.[39]

We start first with estimating accurate projections from passive element signals obtained from experiment of the reciprocal SOS (or slowness) through the object in a maximum likelihood framework.[32, 40] The reciprocal SOS projections are written as TOF values of the path through the object connecting the passive element $\mathbf{r}_{\mathrm{p}}$ with the detector $\mathbf{r}_{\mathrm{i}}$ as:

$$t_f(\mathbf{r}_{\mathrm{i}}, \mathbf{r}_{\mathrm{p}}) = \int_{l(\mathbf{r}_{\mathrm{i}}, \mathbf{r}_{\mathrm{p}}, c)} \frac{1}{c(\mathbf{r})} - \frac{1}{c_0} d\mathbf{r}, \tag{4}$$

where $l(\mathbf{r}_{\mathrm{i}}, \mathbf{r}_{\mathrm{p}}, c)$ is the acoustic ray path connecting the passive element with detector element, and $c_0$ is the speed of sound in the reference medium. This path can be curved, as the ray direction can be bent due to refraction according to Snell's law, when regions of differing SOS are encountered. The path is thus dependent on $c(\mathbf{r})$, which presents a non-linear problem. It can be solved using iterations in a linear way, using a previous estimate $\hat{c}^{(i)}(\mathbf{r})$ to calculate the next estimate $\hat{c}^{(i+1)}(\mathbf{r})$. (See further Reconstruction algorithm for details.)

In the first iteration the ray paths are straight and not dependent on SOS distribution. In the next iterations, the current SOS distribution is used to calculate the curved paths from each passive element position to each detector position. To trace the rays we make use of the Eikonal equation[39]

$$|\nabla t(\mathbf{r})|^2 = \frac{1}{c(\mathbf{r})^2}, \tag{5}$$

where the gradient of the wavefront $\nabla t(\mathbf{r})$, or the TOF field at $\mathbf{r}$, yields the instantaneous vector of a ray passing across the wavefront. To calculate the first arrival time solution to the Eikonal equation, we have employed the fast marching method (FMM).[41] The FMM method is an algorithm that solves the Eikonal equation in a single pass using upwind finite differences. The use of the FMM to account for ray-refraction in UTT has been proposed and investigated by Li *et al.*[33,39,42-44] Specifically, we implement a high accuracy FMM (HAFMM) that applies a second-order approximation to calculate the partial derivatives.[42-44]

For every passive element and every projection, Eq. (5) is solved for $t(\mathbf{r})$, given the estimated SOS map $c(\mathbf{r})$, with the initial condition set for $t(\mathbf{r_p}) = 0$. This results in a TOF map $t(\mathbf{r})$, which contains for every spatial position the shortest TOF through the $c(\mathbf{r})$ map to the starting point $\mathbf{r_p}$, the location of the passive element. From the obtained $t(\mathbf{r})$ map, it is then possible to trace back the ray path belonging to the shortest arrival time via a gradient descent approach through the $t(\mathbf{r})$ map.

*Reconstruction algorithm*

The reciprocal SOS distribution (normalized with slowness of the background medium) is represented on a uniformly sampled rectangular grid. Once the ray paths are known, the problem of reconstructing the slowness distribution is a linear problem. Each projection measurement will be a weighted sum of pixels encountered along the ray path. We sample along segments of each ray path. Every sample of the ray can be expressed as a linear combination of one (using nearest neighbour) or more (using bilinear or bicubic interpolation) pixels around the sample. Subsequent sampling of the ray then gives us the weighted sum of pixels for each projection. Vector $\boldsymbol{x}_c$ holds the normalized slowness values at the grid points:

$$x_{c,k} = \frac{1}{c(\mathbf{r}_\mathrm{k})} - \frac{1}{c_0} \tag{7}$$

If we represent the ray driven discretized measurement model created from the slowness distribution $\boldsymbol{x}_c$ with a projection matrix $\mathbf{H_t}(\boldsymbol{x}_c)$, and the projection measurements with a measurement vector $\mathbf{z_t}$, we have the linear model:

$$\mathbf{z_t} = \mathbf{H}_\mathrm{t}\,(\boldsymbol{x}_c)\boldsymbol{x}_c \tag{8}$$

Starting with an initial guess of the vector $\boldsymbol{x}_c^{(i)}$ we iteratively compute a solution to the regularized cost function[32] using the LSQR method

$$\hat{\boldsymbol{x}}_c^{(i+1)} = \arg\min_{x_c} \left\| \mathbf{z}_\mathrm{t} - \mathbf{H}_\mathrm{t}(\hat{\boldsymbol{x}}_c^{(i)})\boldsymbol{x}_c \right\|^2 + \lambda \mathrm{TV}(\boldsymbol{x}_c) \tag{9}$$

where $\lambda$ is the regularization parameter and TV stands for the Total Variation operator. Some details of the Total Variation (TV) regularization used are provided further in Sec. 3.3, more details will be published in future.[45]

### 3.3: *Reconstruction of SOS compensated light absorption distribution*

At this point we have an SOS distribution (measured and reconstructed as above, or otherwise known) that we use in the process of calculating the integral over iso-TOF contours (Eq. (2)). The SOS grid is resized to cover only the part occupied by the object, an area that encloses the same area as the light absorption map obtained from (simultaneous) PA measurements.

To obtain the TOF map for reconstruction of light absorption, we look again to solving the Eikonal equation (Eq. (5)) using the HAFMM algorithm. This time from the point of view of detector positions, to calculate the integrals over the iso-TOF contours of Eq. 2. The procedure is repeated for all detector positions, resulting in $\boldsymbol{n}$ TOF maps. Since the grids of the optical absorption map and the TOF maps are not necessarily the same, we use bicubic interpolation to obtain the TOF values at off grid points.

Continuing from Sec. 3.1, Eq. 3 is solved for $\boldsymbol{x}_A$ iteratively, because the number of equations and unknowns is large. Anastasio *et al*[27] used an EM algorithm, Jin and Wang [28] used LSQR, Jin Zhang *et al*[46] used a truncated conjugate gradient (TCG) method with a roughness penalty, and Zhang and Wang[25] used a modified filtered backprojection applied over iso-TOF curves. We preferred using the LSQR, since the EM method requires more iterations.[47] Further, LSQR has more favorable numerical properties compared with the TCG method,[48] while FBP is an approximation that results in lower accuracy.

When the projection matrix in Eq. 3 is ill-conditioned, the inversion can be undefined, since a part of the singular values can be zero or close to zero. Regularization is resorted to, in order to stabilize the inversion, which means that prior information about the solution is added to the problem.[49] While the most general form of regularization is the Tikhonov regularization[50] which uses an $L^2$ norm, (which penalizes the higher frequencies according to a quadratic criterion) the use of a regularizer based on an $L^1$ norm[47] provided better results with our data. This Total Variation (TV) regularization, in addition to providing a smoothing effect also preserves edge information in the reconstruction. The TV regularized solution is calculated from:

$$\hat{\boldsymbol{x}}_A = \arg\min_{\boldsymbol{x}_A} \left\| \mathbf{z}_{\mathrm{p}} - \mathbf{H}_{\mathrm{dt}} \mathbf{H}_{\mathrm{TOF}} \boldsymbol{x}_A \right\|^2 + \lambda \mathrm{TV}(\boldsymbol{x}_A) \tag{10}$$

where $\lambda$ the regularization parameter controls the smoothness of the solution, and $\mathrm{TV}(\boldsymbol{x}_A) = \sum_i \sqrt{(\partial_x x_i)^2 + (\partial_y x_i)^2 + \beta^2}$. The constant $\beta > 0$ offers computational advantages, such as differentiability when the gradient approaches zero.[51] Effectively the $\beta$ constant allows a smooth transition from aquadratic, $L^2$ cost function, for small gradients to a linear, $L^1$ cost function for large gradients. (Similar considerations were applied in arriving at Eq. 9 for the SOS reconstruction.)

## 4: Results and discussions

### 4.1: *Numerical phantom*

*Reconstruction of SOS*

To investigate the impact of curved rays on the reconstruction of SOS, we applied both the straight-line approach and the present iterative bent ray approach to the data. Figure 3(a) shows that with straight ray paths, the shapes and sizes of the inhomogeneities are distorted, and many severe artifacts are present. The result from the iterative approach in a 15$^{th}$ iteration (Fig. 3(b)), show excellent reconstruction with correct retrieval of sizes, shapes and SOS values, and a complete removal of artifacts.

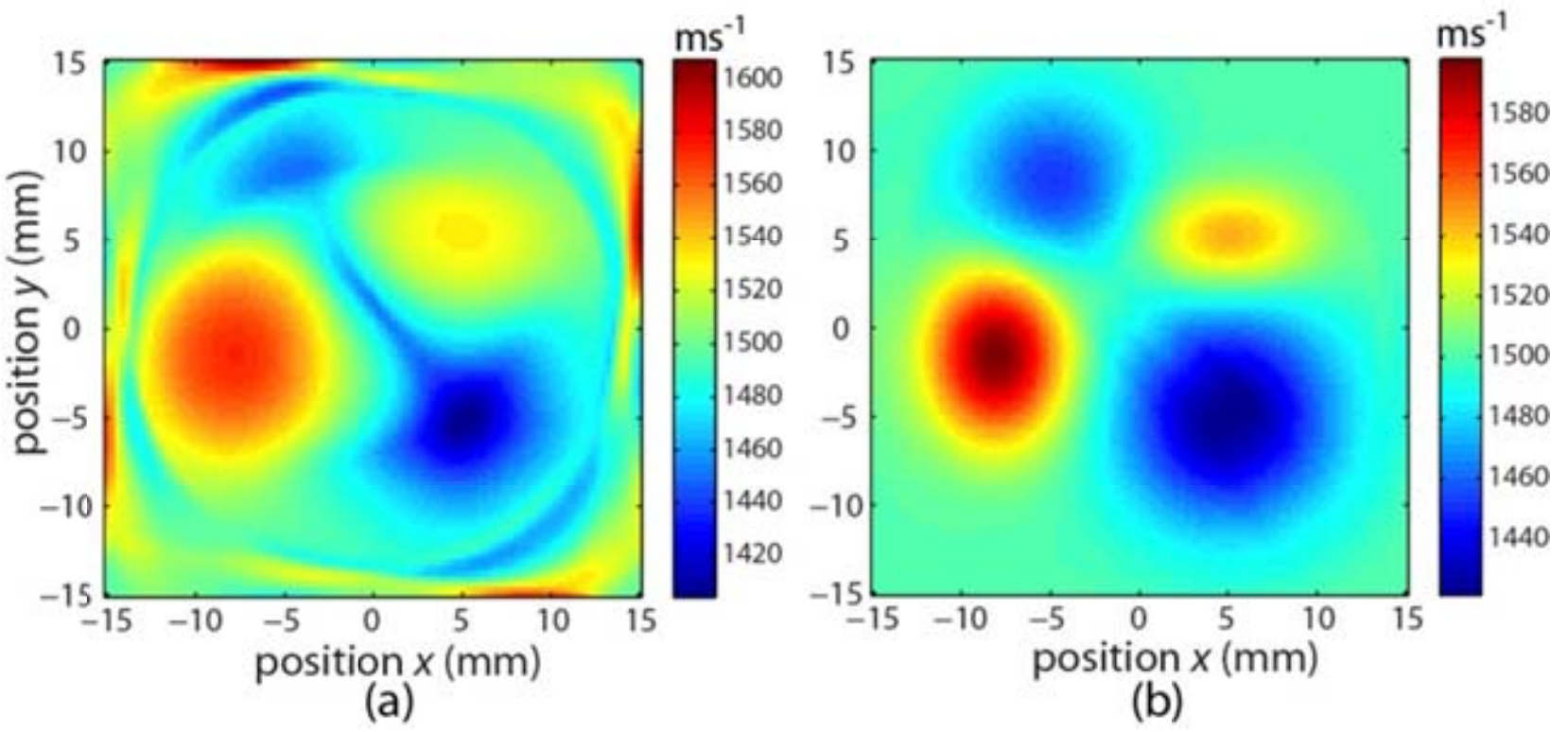

**FIG3**.SOS reconstruction results of the numerical phantom, (a) with the assumption of straight ray acoustic propagation, (b) using 15 iterations of the present approach that accounts for curved rays using the HAFMM.

*PA reconstruction*

The PA image reconstructed using a single SOS is shown in Fig. 4(a), with a horizontal profile plot through the center of the image in Fig. 4(c). The background SOS value of 1500 ms$^{-1}$ was used. Blurring artifacts resulting from ignoring SOS inhomogeneities are evident. The distortion in the smallest structure (above the center) is most noticeable in the image.

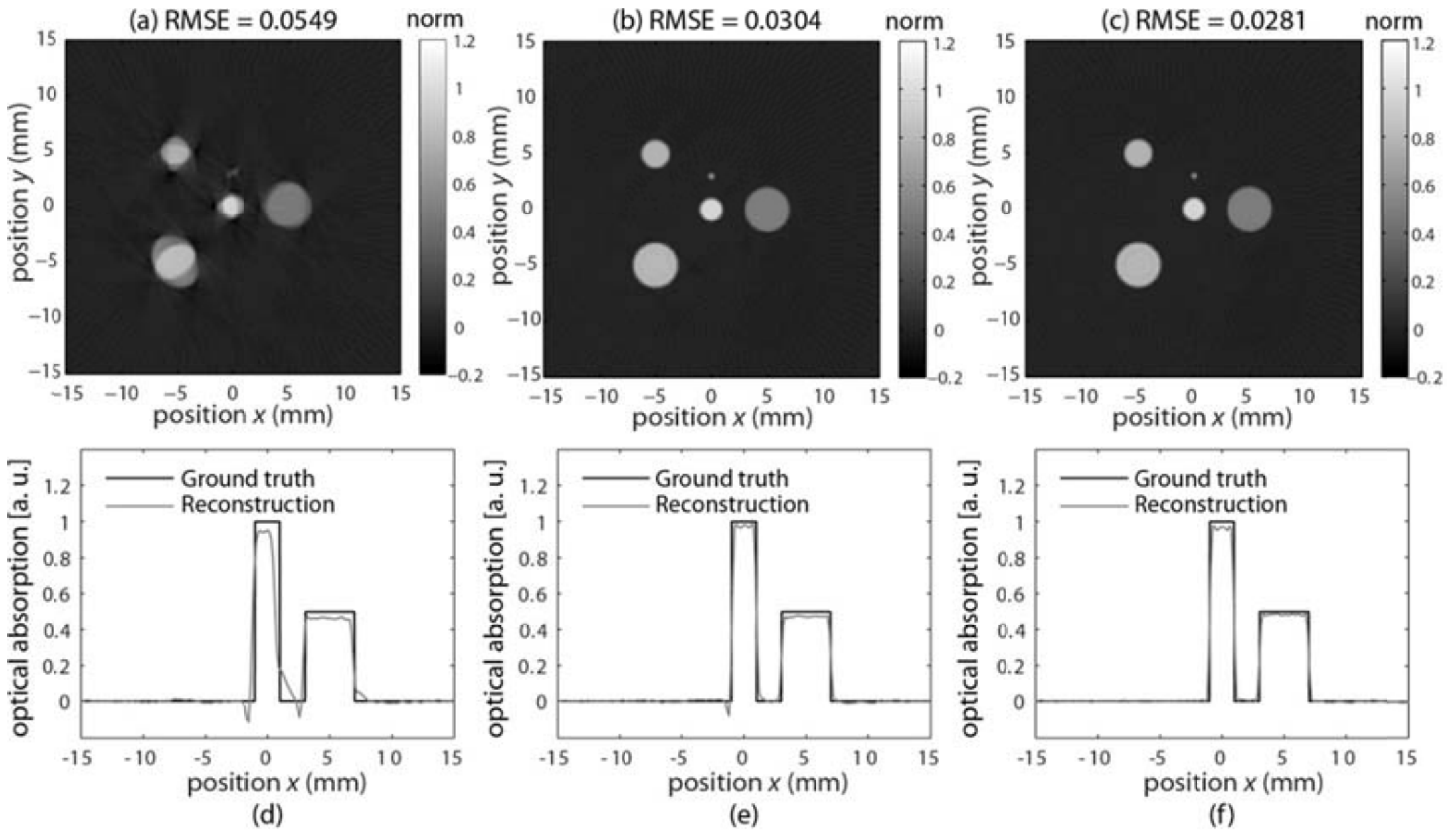

**FIG4**.Optical absorption tomograms of the numerical phantom using, (a) reconstruction assuming a uniform SOS of 1500 $ms^{-1}$, (b) reconstruction using straight rays through the SOS distribution, (c) the present SOS compensation method that uses refracted rays. Horizontal profile plots through the center of the image reconstructed with (d) uniform SOS, (e) straight rays through the SOS distribution, and (f) with the present method.

Figures 4(b) and 4(e) show the result obtained using straight-ray propagation. Here iso-TOF contours were calculated by tracing straight rays through the SOS map. The result is considerably improved compared to the previous case with a complete removal of artifacts. Finally Fig. 4(c) shows the result using the present method that uses SOS distributions in the reconstruction, where iso-TOF curves are computed using bent rays in the HAFMM method. The results are artifact-free as in the straight-ray approach, but with a lower RMSE error. The improvement obtained using the present approach can best be appreciated by an examination of Fig 4(f): the horizontal profile shows hi-fidelity reconstruction compared with ground truth, better than that obtained using the straight-ray approach.

**4.2*: Leaf Skeleton imaging***

Figure 5(a) shows the PA image of the leaf skeleton reconstructed assuming a uniform SOS for the entire imaging area. The image has pixel sizes of 100x100 $\mu$m. We chose $\lambda$=1 and $\beta$=28 in the reconstruction according to Eq. 10. An optimized value of SOS was chosen, taking into account the SOSs of the various media in the region. We calculated a weighted average of 1506 $ms^{-1}$ for the imaging area due to the relative presence of water (1492 $ms^{-1}$), PVA (1575 $ms^{-1}$) and agar (1498 $ms^{-1}$). Double-line artifacts are evident in the reconstruction of the veins due to the presence of the symmetrical SOS inhomogeneity around the sample. To illustrate the effect more clearly, a close up of the smallest venules (from within the white dotted square in Fig. 5(a)) is shown in Fig. 5(d).

It is clear that assuming a uniform SOS value for the imaging area introduces errors in the registration of the correct spatial location of the acoustic sources, resulting in degradation in the resolution and contrast of the reconstructed images. The PA image reconstructed along iso-TOF contours calculated from the SOS map is shown in Fig. 5(b) when using straight-ray propagation; and in Fig. 5(c) using the present curved-ray approach. In both cases the images are superior to the uncorrected case, showing the vein morphology as expected, with the complete absence of double-line artifacts and minimized blurring. Careful examination of the corresponding zoomed in images (Figs. 5(e) and (f)) shows that our curved-ray based reconstruction gives slightly better results than the straight-ray method: the midvein is compacter in Fig. 5(f) than in Fig 5(e), and the smallest partly connected blobs of the quaternary venules in Fig. 5(e) are better interconnected in Fig. 5(f). The smallest venules measured using an optical microscope to be 30 $\mu$m in diameter are reconstructed with a size of 100 $\mu$m, which is the resolution limit of the system.[32]

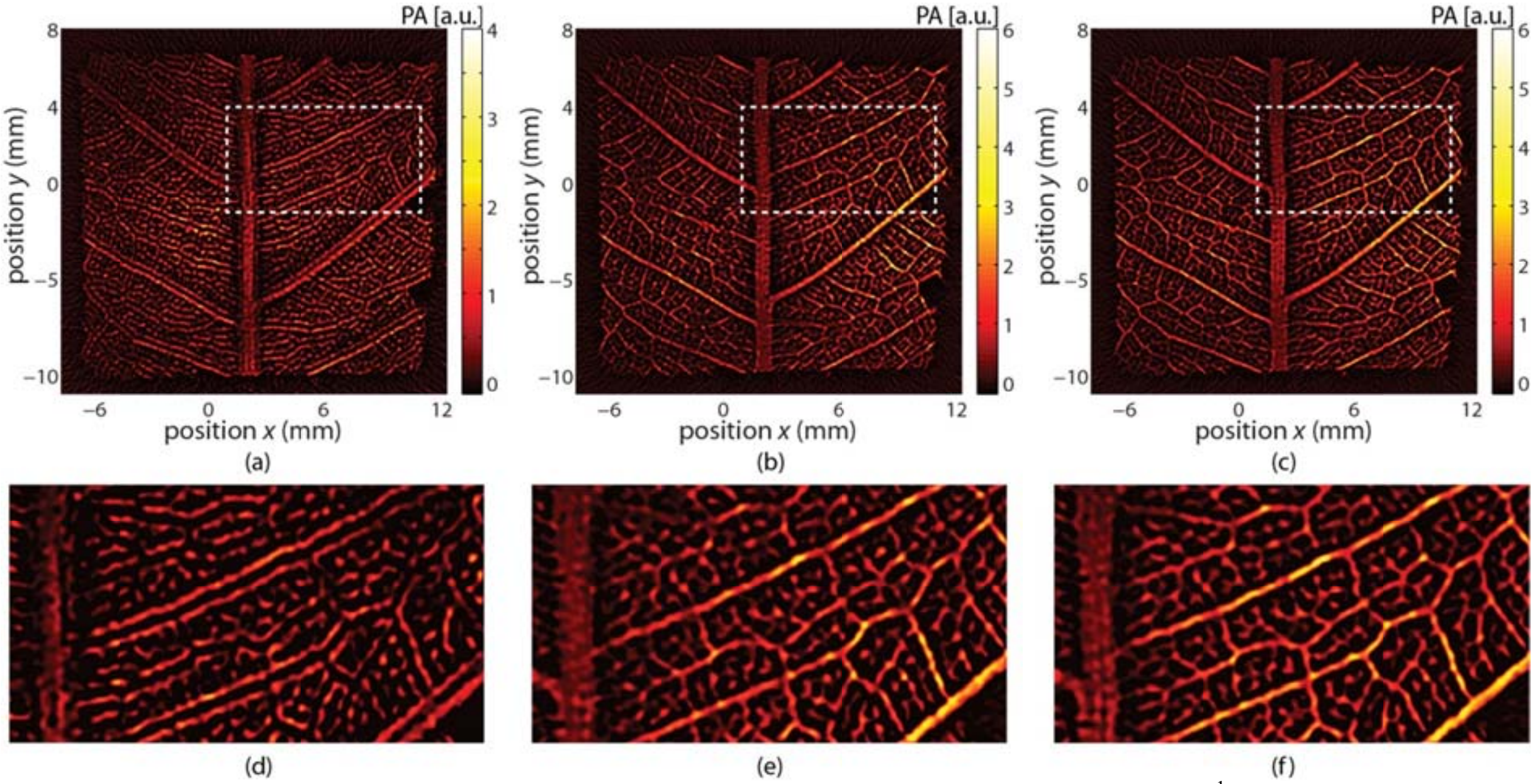

**FIG 5**: PA image of the leaf skeleton reconstructed with (a) a uniform SOS of 1506 ms$^{-1}$, the weighted average of SOS of all media (water, agar and PVA) considering their relative presence, (b) the SOS map using the straight-ray propagation approach, and (c) the SOS map using the curved-ray approach. In (d), (e) and (f) are the corresponding zoomed-in views of the veins (within the boxes indicated with dashed lines above).

### 4.3: *Kidney imaging*

Figure 6(a) shows the SOS tomogram reconstructed from the acquired passive element data of the imaging area using the bent ray approach. The SOS image grid has pixel sizes of 400x400 $\mu$m. We chose $\lambda$=20 and $\beta$=0.04 in the reconstruction according to Eq. 10. The SOS recovered from the kidney region is 1535 ms$^{-1}$, which is close to the value reported in literature.[52] The water region shows an SOS value of 1492 ms$^{-1}$ which matches the values for water at 22$^{0}$ C.[53] The embedding agar also appears to be visible with a value of 1498 ms$^{-1}$. Figure 6(b) shows the PA image of the kidney reconstructed using an SOS value of 1540 ms$^{-1}$, the commonly used value for soft tissue. The blood vessels are depicted with low contrast and blurring, with some double-line artifacts.

When we reconstruct the PA image with pixel sizes of 100x100 $\mu$m ($\lambda$=1 and $\beta$=28) using the TOF map generated from the reconstructed SOS tomogram using the bent-ray approach, the image shows significant improvements (Fig. 6(c)). The blood vessels are sharp without blurring and the artifacts seen in Fig. 6(b) are minimized. Further, features such as interlobular, interlobar, segmental and renal blood vessels may now be recognized. In this particular case, there appeared to be only marginal changes between the straight-ray (not shown) and bent-ray approaches for SOS compensation.

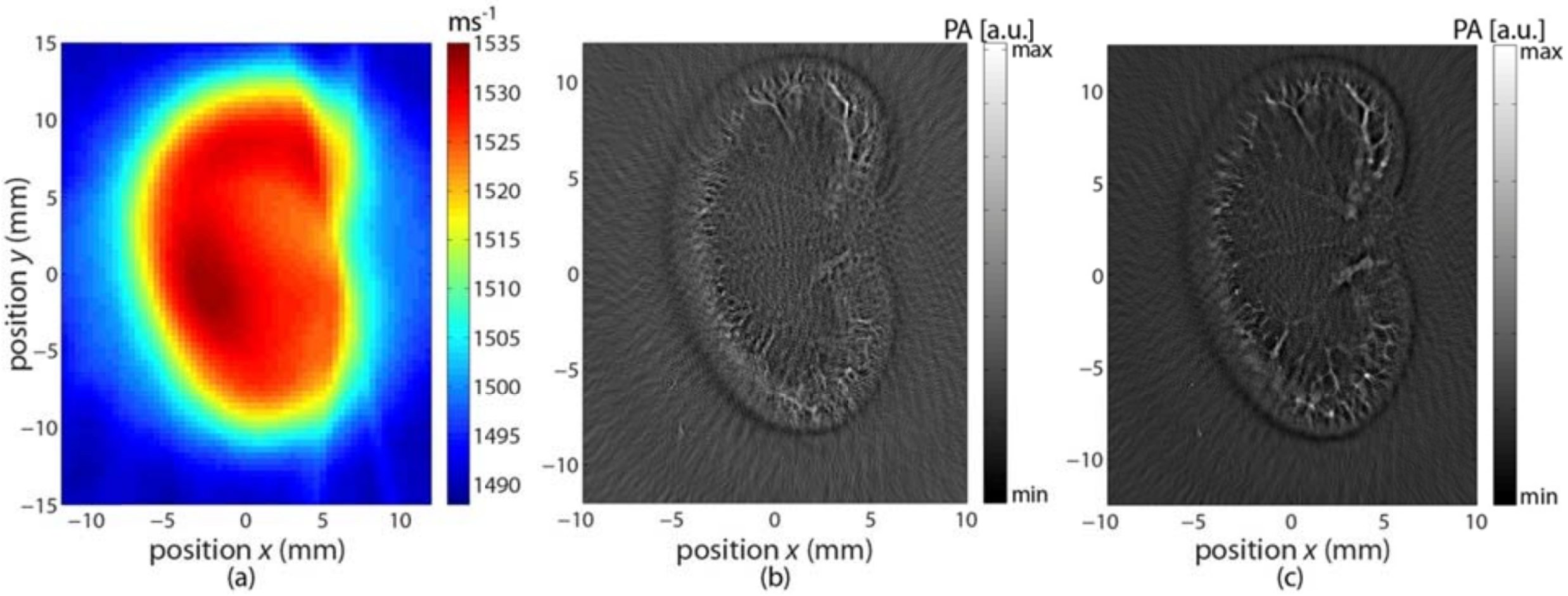

**FIG 6** (a) SOS map reconstructed from the acquired passive-element measurement of the imaging area, (b) PA image reconstructed with a uniform SOS of 1540 ms$^{-1}$, (c) SOS compensated PA image using the SOS values from (a).

**Discussion**

We showed numerically and experimentally 1) that using a 2-D iterative inversion algorithm which takes into account ray refraction by solving the Eikonal equation, we can retrieve accurate SOS images in a hybrid PER-PACT imager; 2) that using a novel iterative 2-D PA reconstruction algorithm which also accounts for ray-refraction, we can use the SOS map to compensate PA images for SOS inhomogeneities. The end result is accurate images that have considerably minimized artifacts.

Obviously, the most accurate reconstructions would consider the full wave equation description of ultrasound propagation, whose solutions are computationally very burdensome. In the applications envisaged for our methods, namely small-animal imaging or potentially in human breast imaging, we expect smoothly varying refraction from various soft-tissue types to be dominant over other wave effects such as diffraction. Further, the sizes of the SOS or refractive inhomogeneities will be larger than the ultrasound wavelengths. In such cases, geometrical acoustics, where the energy travels along rays, is a good approximation to the full wave equation while being considerably simpler to compute. Within the geometrical acoustics approximation, the use of bent-rays is more accurate than straight rays. This contributes to superior images compared with using straight rays as we have seen in simulations (Figure 4), and in the leaf experiments (Figure 5). For the kidney experiments the differences between the straight-ray and bent-ray reconstructions are not perceptible possibly due to the absence of a large inhomogeneity as in in the case of the PVA ring in the leaf experiments.

For the PA reconstructions of the leaf and kidney data, we set the maximum number of iterations to 20; convergence is achieved under this. The number of measurement signals for a slice is 576 (18 projections x 32 elements), and the number of pixels usually chosen for the reconstruction is 250x250 pixels (to include more than the object) with pixel sizes of 100x100 $\mu$m. We have an SOS image grid covering 100x100 pixels with pixel sizes of 400x400 $\mu$m. To illustrate the computational times required with the above numbers: the leaf data reconstruction as presented took 4208.586 s using the bent-ray approach, and a slightly less 3330.266 s when the straight-ray approach was used. These times were obtained using an Intel(R) Core(TM) i7-2600 CPU @ 3.4 GHz and 8.00 GB RAM running 64-bit (win64) Matlab R2010b with multithreaded mex functions to calculate the photoacoustic projection integral.

While we will continue to test the feasibility of our hybrid tomography PER-PACT approach and reconstruction/compensation algorithms in more complicated phantoms leading to *in vivo* studies, it should be acknowledged that our approach is only an approximation to 2-D. The reason for this is that even though the PER-PACT instrumentation has been designed to acquire data from a 2-D slice, the elevation dimension is not actually reduced to zero in this or any other practical implementation of an imager. This leads to non-uniform resolutions (125 $\mu$m in the imaging plane, compared with 1 mm in the elevation plane) which can cause artifacts if samples are not appropriately chosen. Extension to 3-D instrumentation and models will improve the accuracy of reconstruction,[54] however will lead to severely longer computation times than presented above. For 3-D reconstruction the number of unknowns (voxels in the imaging volume) and the number of measurements are increased. If in a 2-D situation an image comprises $N$x$N$ pixels, in 3-D assuming uniform resolution the image would require $N$x$N$x$N$ voxels. If $M$ additional measurements are required, the total reconstruction time will increase by a factor $N$x$M$. Thus, if 3-D iterative image reconstruction algorithms are used they would require implementation on specialized hardware in the form of highly parallelized processing architectures as in GPUs (Graphics Processing Units).[54]

## Concluding remarks

Most PA studies and reconstruction algorithms are based on the assumption of tissue homogeneity which justifies the use of a single SOS value for the region of interest. We have demonstrated numerically and experimentally that such an assumption in the presence of SOS heterogeneities leads to inaccurate reconstruction of PA tomograms. We have also demonstrated that the use of our PER-PACT approach,[32] allowing simultaneous acquisition of PA and SOS information, with the use of suitable reconstruction algorithms removes degradation in PA image quality from SOS heterogeneities. The iterative reconstruction algorithm for retrieving SOS images, using the Eikonal equation to model refractive effects in the forward projection, yields accurate SOS images in the PER-PACT setting. Further, we have introduced a novel iterative PA reconstruction algorithm utilizing SOS information, which for the first time accounts for ray-refraction also using the Eikonal equation. This approach is successful in obtaining artifact-free highly accurate PA tomograms. Both algorithms use a high accuracy fast marching method (HAFMM) for computation of TOF values at each pixel in the reconstruction grid. In addition to improving image quality in PA, knowledge of the spatial distribution of SOS can also potentially provide diagnostic value. It is known from clinical studies in UTT that the SOS (and acoustic attenuation - AA) depend on disease state, and SOS (and AA) tomograms are being evaluated for detection and diagnosis of breast disease.[29] Cancerous tissue has an average SOS of 1559 $ms^{-1}$, which is different from that in surrounding fat (≈ 1470 $ms^{-1}$) and glandular tissue (≈ 1515 $ms^{-1}$).

## Acknowledgements

We thank Bart Koornwinder, Steffen Resink, and Johan van Hespen from the Biomedical Photonic Imaging group (BMPI), University of Twente for their considerable contributions. This research was funded by the MIRA Institute for Biomedical Technology and Technical Medicine of the University of Twente via the NIMTIK program; and by AgentschapNL through the projects IPD067771 (PRESMITT) and IPD083374 (HYMPACT) in the theme IOP Photonic Devices.

**LIST OF FIGURES:**

**FIG 1**: Schematic of the experimental setup, with multiple side illumination

**FIG 2**: Overview of the phantoms used. (a) Numerical phantom showing speed-of-sound (SOS) distribution in $ms^{-1}$, (b) same phantom showing the distribution of optical absorption in arbitrary units, (c) photograph of the leaf skeleton piece embedded in an agar cylinder with an enclosing high velocity PVA cylinder. The values of the SOS of various media are also marked. (b) Resected rat kidney embedded in an agar cylinder.

**FIG 3**.SOS reconstruction results of the numerical phantom, (a) with the assumption of straight ray acoustic propagation, (b) using 15 iterations of the present approach that accounts for curved rays using the HAFMM.

**FIG 4**.Optical absorption tomograms of the numerical phantom using, (a) reconstruction assuming a uniform SOS of 1500 $ms^{-1}$, (b) reconstruction using straight rays through the SOS distribution, (c) the present SOS compensation method that uses refracted rays. Horizontal profile plots through the center of the image reconstructed with (d) uniform SOS, (e) straight rays through the SOS distribution, and (f) with the present method.

**FIG 5**: PA image of the leaf skeleton (a) reconstructed with a uniform SOS of 1506 $ms^{-1}$, the weighted average of SOS of all media (water, agar and PVA) considering their relative presence, and (b) reconstructed using the known SOS distribution. Zoomed-in views of the veins reconstructed with (c) a uniform weighted-average SOS of 1506 $ms^{-1}$ and (f) complete SOS map.

**FIG 6** (a) SOS map reconstructed from the acquired passive-element measurement of the imaging area, (b) PA image reconstructed with a uniform SOS of 1540 $ms^{-1}$, (c) SOS compensated PA image using the SOS values from (a).